\documentclass[fleqn,twoside]{article}
\usepackage{amsmath}
\usepackage{espcrc2}
\usepackage{amssymb}
\usepackage{psfrag}
\usepackage{graphicx}
\usepackage{latexsym,amsfonts}

\begin{document}

\title{Isospin breaking in $\pi $N scattering at threshold by radiative processes}
 
\author{ T.E.O.~Ericson $^{a}$\thanks{Corresponding author.  \mbox{\it
      E-mail }: torleif.ericson@cern.ch} ~and~ A. N.
  Ivanov${^b}$\,\thanks{\mbox{\it E--mail}: ivanov@kph.tuwien.ac.at .
    Permanent Address: State Polytechnic University, Department of
    Nuclear Physics, 195251 St. Petersburg,
    Russian Federation}\\
  \addressmark{$^a$ Theory Division, Physics Department, CERN, 
CH-1211 Geneva 23, Switzerland, \\
    $^b$ Atomic Institute of the Austrian Universities, Vienna
    University of Technology, \\ Wiedner Hauptstrasse 8-10, A-1040 Wien and 
    SMI of the Austrian Academy of Sciences, Boltzmanngasse 3, A-1090,
    Wien}}

\begin{abstract}
  We investigate the dispersive contribution by radiative processes such as
  $\pi^-p \to n\gamma$ and $\pi^- p \to \Delta \gamma$ to the 
  $\pi N$ scattering lengths of charged pions in the heavy baryon limit. 
They give  a large isospin
  violating contribution in the corresponding isoscalar scattering length, but only
  a small violation in the isovector one.  These terms contribute $6.3(3)$\% 
 to the 1s level shift  of pionic hydrogen and give a chiral constant $F_{\pi }^2f_1=-25.8(8) $ MeV.

 PACS: 11.10.Ef, 11.55.Ds, 13.75.Gx, 36.10.Gv
\end{abstract}

\maketitle

\section{Introduction}
The energy shift and width in the $\pi ^-p$ atom as compared to the
purely electromagnetic bound state energies  have recently been
measured to the remarkable precision of $\pm 0.2\,\%$ and $\pm 1\,\%$,
respectively \cite{PSI1} (see also \cite{PSI2}). These quantities are
proportional to the corresponding real and imaginary part of the
threshold $\pi ^-p$ scattering amplitude (the scattering length)
\cite{DT54,TT61}, which therefore are determined to the same precision
provided electromagnetic corrections on the several \% level can be
properly understood and accounted for.

In the absence of the external Coulomb field, the real and imaginary
parts of the S--wave amplitude $f^{\pi^-p}(0)$ at threshold ($q=0$,
where $q$ is the relative momentum of the $\pi^-p$ pair )  are equal~to
\begin{eqnarray}\label{label1} {\cal R}e\,f^{\pi^-p}(0)
  \,=\,a_{cc}\;,\;{\cal I}m\,f^{\pi^-p}(0)
  \,=\,q_{\pi^0 n}\,a^2_{nc}.
\end{eqnarray}  
Here $a_{cc} = a_{\pi^-p \to \pi^-p}$ and $a_{nc} = a_{\pi^-p \to
  \pi^0 n}$ are the S--wave scattering lengths of the reactions
$\pi^-p \to \pi^-p$ (the charged channel (cc)) and of $\pi^-p \to
\pi^0 n$ (the charge exchange one), respectively, and $q_{\pi^0 n} =
28.040\,{\rm MeV}$ is the relative momentum of the $\pi^0 n$
system\footnote{We have removed the physical contribution to the
  imaginary part from the radiative channel $\pi ^-p\to \gamma n$,
  which is accurately known from the Panofsky ratio
  $P\,=\,\sigma(\pi^-p \to \pi^0n)/\sigma(\pi^-p \to \gamma n) =
  1.546(10) $\cite{JS77}.}.

One has then, in principle, 
an exceptional source for the
hadronic scattering lengths which can be directly compared to the
large body of phenomenological $\pi $N phase shift data as well as to
predictions of Chiral Perturbation Theory (ChPT). 
The electromagnetic corrections to the observed
energy level shift
are of two types: a) iterated ladders of photon
 exchange with  the intermediate $\pi $N state remaining in its ground state
and b) corrections corresponding to inelastic intermediate states.
The first class of corrections has recently been investigated
\cite{ERI04}.  These moderate corrections were found to correspond to well
understood physical effects to a precision commensurate with the present
experimental one.  While such effects violate isospin, 
it is not usual to consider them as genuine isospin breaking. 
 The second group of e. m.  corrections are intrinsic to the
 scattering processes. They produce genuine isospin
breaking together with the isospin breaking from the strong
interaction itself.  In order to reduce the data to pure hadronic
interactions one must therefore  control these intrinsic e.~m.
 contributions to sufficient accuracy
and  understand their physics.
\footnote{ Another use of the relation (1) is the description of the
  energy shift in pionic hydrogen in terms of ChPT \cite{JG02}. In
  this case terms of different origin are not separated.  The e.~m.
  contributions include implicitly the Coulomb terms}.

Here we investigate this second class of electromagnetic corrections.
There are excellent reasons to believe that such electromagnetic processes
contribute substantially to the $\pi ^-$p scattering length.  In fact,
the dominant contribution (65\%) to the 1s width of pionic hydrogen is
the radiative capture by the electric dipole transition $\pi ^-p
\rightarrow n \gamma $, the so-called Kroll-Ruderman process~\cite{KR54} (Fig.1). The corresponding radiative width is no less
than 8\% of the strong interaction shift as observed in the
 Panofsky ratio~\cite{JS77}. This indicates that the  corresponding  dispersive
 term may contribute  4\% and
even more to the energy shift in pionic hydrogen.  This is indeed the
case as we will show.

Since the Kroll-Ruderman contribution is generated from the pion
p-wave part of the nucleon pole term, one must inevitably consider
also other important aspects of the well understood p-wave $\pi N$
physics.  Here the $\Delta $ isobar and the $N\Delta $ mass difference
 are central
features with the $\Delta $ pole as important as the nucleon
one \cite{TE88}. We therefore include it on an equal footing with the nucleon.
Since the main purpose of the present paper is to clarify the 
physical mechanisms of e.~m.  isospin breaking in the $\pi $N
scattering lengths in simple terms, we rely on the heavy baryon limit.
This is in accordance with p-wave $\pi N$ phenomenology for which this
approximation describes well the main aspects of the interaction. It
will become apparent that the bulk of the isospin breaking occurs
already in the limit of a vanishing pion mass. The finite pion mass
introduces small, but characteristic, additional terms.  

The heavy baryon limit, used in chiral perturbation expansion as well,
and the threshold condition lead to great simplifications of the
problem. The relevant contribution becomes that of an electric dipole
process (E1) due to transition radiation by the absorption or
emission of the charged pion.  In particular, there is no coupling
 to the baryon convection current nor to its magnetic moment in
this limit, while the threshold condition suppresses radiation
produced by changes in the pion convection current.

The leading e.~m. isospin breaking effects in low energy $\pi N$ scattering have
previously been discussed in particular using Heavy Baryon ChPT [see Refs. \cite{FET01} and
references therein] as
well as using a heavy quark model \cite{VL01}.  Such isospin breaking is 
implicit in several theoretical
studies of the contributions to the 1s energy shift of the $\pi ^-p$
atom to leading \cite{VL00} and next to leading order power counting in an
effective field theory(EFT)  of QCD+QED  \cite{JG02}.

The paper is organized as follows. In Section 2 we derive the general
expression for the dispersive e.~m. contributions to the S--wave
amplitude of elastic $\pi^-p$ scattering at threshold, saturated by
intermediate $X \gamma$ states related to the reactions the $\pi^-p
\to X \gamma$, where $X$ is a hadronic state (see Fig.~1).
\begin{figure}
{\hspace{-1in}\psfrag{pm}{$\pi = \pi^{\pm}$}
\psfrag{p}{$ N = p, n$}
\psfrag{g}{$\gamma$}
\psfrag{X}{$X = N, \Delta, \ldots$} \centering
\includegraphics[height=0.15\textheight]{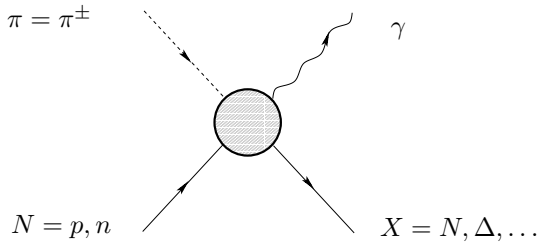}
\caption{Diagram of  $\pi N \to X \gamma$
  reactions with one photon in the intermediate state generating a 
 contributiion to low--energy elastic $\pi N$ scattering. }}
\end{figure}
In Section 3 we apply the results specifically to  the calculation
of the dispersive contributions from the reactions $\pi^- p \to n \gamma$ and
$\pi^-p \to \Delta \gamma$, respectively, and generalize this to
the elastic $\pi^c N$ threshold amplitude for any charged pion. In Section 4
we discuss the individual contributions to the isospin breaking under
different assumptions and discuss the relation of our results to other
investigations. In the Conclusion we summarize the results.  

\section{One--photon exchange contributions to the
  elastic $\pi^-p$ threshold amplitude }

For concreteness we will illustrate these contributions for the case
of $\pi ^-$p elastic scattering at threshold, but the argument is
nearly identical for any elastic $\pi N$ channel with a charged pion.
 The S--wave amplitude of the $\pi^-p $ scattering caused by
the  channels $\pi^- p \to X \gamma \to \pi^- p $
with one intermediate photon
(see Fig.~1) is defined by  
\begin{eqnarray}\label{label2}
  \hspace{-0.2in}&&(1+{m_{\pi }\over M_N})\, \delta f^{(\gamma )} = \frac{1}{8\pi M_N}
  \nonumber \times \\
  \hspace{-0.2in}&& \lim_{TV
 \to
    \infty} \frac{\langle
    \pi^-(\vec{0}\,)p(\vec{0},\sigma_p)|
{\cal T}|\pi^-(\vec{0}\,)p(\vec{0},\sigma _p)\rangle }{TV},
\end{eqnarray}
where $TV= (2\pi)^4\delta^{(4)}(0)$ is a 4--dimensional volume and $M_N$ the nucleon mass.
The ${\cal T}$
 matrix is related to the $S$ matrix by $S = 1 +i{\cal T}$,
where $S$ is defined by
\begin{eqnarray}\label{label3}
S = {\rm T}\exp\,\Big\{-\,i\int
d^4x\,\sqrt{4\pi}\,e\,J_{\mu}(x) {\cal A}^{\mu}(x)\Big\}.
\end{eqnarray}
Here $T$ is the time--ordering operator, $J_{\mu}(x)$  the hadronic
electromagnetic current \cite{SA68,ME72} and ${\cal
  A}^{\mu}(x)$  the electromagnetic potential.

 Integrating over photon degrees of freedom 
\begin{eqnarray}\label{label4}
 S = \exp\,\Big\{\,i\int d^4x\,{\cal L}^{\rm
      em}_{\rm eff}(x)\Big\}.
\end{eqnarray}
The effective Lagrangian ${\cal L}^{\rm em}_{\rm eff}(x)$ is
given by
\begin{eqnarray}\label{label5}
 {\cal L}^{\rm em}_{\rm eff}(x) &=& 2\pi \alpha\, i\int
d^4y\,{\rm T}(J_{\mu}(x)J_{\nu}(y))\nonumber\\
 &\times&\langle 0|{\rm T}({\cal A}^{\mu}(x)
{\cal A}^{\nu}(y))|0\rangle
\end{eqnarray}
with
\begin{eqnarray*}
  \hspace{-0.3in}&&\langle 0|{\rm T}({\cal A}^{\mu}(x)
  {\cal A}^{\nu}(y))|0\rangle = \theta(x^0 - y^0)\int \frac{d^3p}{(2\pi)^3 
    2|\vec{p}\,|}\nonumber\\
  \hspace{-0.3in}&&\times\,e^{\textstyle -ip\cdot (x - y)}\sum_{\lambda} 
  e^{*\mu}(\vec{p},\lambda)e^{\nu}(\vec{p},\lambda) + (x \leftrightarrow y).
\end{eqnarray*}
Here $e^{\mu}(\vec{p},\lambda) = (0, \vec{e}\,(\vec{p},\lambda))$ is
the polarization vector of the photon  in the Coulomb gauge. To
lowest order in the fine structure constant $\alpha \simeq 1/137.036 $, the threshold contribution is
\begin{eqnarray}\label{label6}
  \hspace{-0.3in}&&\Big(1+ \frac{m_{\pi }}{ M_N}\Big)\, \delta f^{(\gamma)} = \frac{1}{8\pi M_N}\times \nonumber \\
  \hspace{-0.3in}&&\,\langle \pi^-(\vec{0})p(\vec{0},\sigma_p)|
  {\cal L}^{\rm
    em}_{\rm eff}(0)|\pi^-(\vec{0})p(\vec{0},\sigma_p)\rangle. 
\end{eqnarray}
Substituting Eq.~(\ref{label5}) into Eq.~(\ref{label6}) and averaging
over polarizations of the proton we get
\begin{eqnarray}\label{label7}
  \hspace{-0.3in} &&\Big(1+ \frac{m_{\pi }}{M_N}\Big)\, \delta f^{(\gamma)} = \frac{\alpha}{8\pi M_N}\int\frac{d^3p}{(2\pi)^2 2 |\vec{p}|}\nonumber \times \\
  \hspace{-0.3in} &&\, W_{\mu\nu}(p, k,
  q)\sum_{\lambda }e^{*\mu}(\vec{p},\lambda)
  e^{\nu}(\vec{p},\lambda)\Big|_{ \vec{k} = \vec{q} = 0},
\end{eqnarray}
where we have introduced the structure tensor $W_{\mu\nu}(p, k,q)$ as
follows:
\begin{eqnarray}\label{label8}
  \hspace{-0.3in}  &&W_{\mu\nu}(p,k,q) = i\int d^4x\,\theta( x^0)\,
  e^{\textstyle\,- ip\cdot x}\nonumber\\
  \hspace{-0.3in} &&\times\,\frac{1}{2}\sum_{\sigma_p = \pm 1/2} \langle
  \pi^-(\vec{k}\,)p(-\vec{k},\sigma_p)|\left(
  J_{\mu}(x)J_{\nu}(0) \right.\nonumber\\
   \hspace{-0.3in} && + \,\left.
  J_{\nu}(0)J_{\mu}(-x)\right)
  |\pi^-(\vec{q}\,)p(-\vec{q},\sigma_p)\rangle.
\end{eqnarray} 
For the calculation of $W_{\mu\nu}(p,k,q)$ we insert the complete set
of the intermediate hadronic states
\begin{eqnarray*}
\sum_X |X\rangle \langle X| = 1,
\end{eqnarray*} 
where $|X\rangle$ is an arbitrary hadron state with baryon number $B_X
= 1$. 

We now eliminate the pion field from the current matrix element
$\langle X|\,J_{\mu }|N\pi^{\pm }\rangle $ following the standard
reduction technique in favour of the hadronic axial current $\langle
X|J^{(\pm)}_{5 \mu}|N \rangle $, where the isospin label $J^{(\pm)}_{5
  \mu }(x)= J^{(1)}_{5 \mu }(x) \pm i\,J^{(2)}_{5 \mu }(x)$.  This is
achieved, for example, using  PCAC and minimal electromagnetic
coupling together with soft--pion assumptions \cite{SA68,ME72}.
One finds for the case of $\pi ^-p\to X \gamma $:
\begin{eqnarray}\label{label9}
  \hspace{-0.3in}&&\langle X|J_{\mu}(0)|\pi^-(\vec{k}\,) 
p(-\vec{k},\sigma_p)
  \rangle =
  \nonumber\\
  \hspace{-0.3in}&&= -\,\frac{i}{\sqrt{2}F_{\pi}}\langle X| 
J^{(-)}_{5\mu}(0)|p(\vec{0},\sigma_p)
  \rangle.
\end{eqnarray}
Here $F_{\pi} = 92.4(3)\,{\rm MeV}$ is the pion decay constant.
  In the soft--pion limit this gives the Kroll--Ruderman
theorem \cite{KR54} (see e. g. \cite{TE88,ME72}).

Substitution into Eq. (\ref{label7}) and integration gives the
structure tensor in the form
\begin{eqnarray}\label{label10}
\hspace{-0.3in}&&W_{\mu\nu}(p,k,q) =
\frac{1}{F^2_{\pi}}\sum_X\Big[\frac{(2\pi)^3\delta^{(3)}(\vec{p} -
\vec{Q}_X)}{ E_X + |\vec{p}\,| - m_{\pi} - M_N - i0}\nonumber\\ 
\hspace{-0.3in}&&\times\frac{1}{2}\sum_{\sigma_p = \pm 1/2}\langle
p(\vec{0},\sigma_p)|J^{+}_{5\mu}(0)|X\rangle \langle X|
J^{-}_{5\nu}(0)| p(\vec{0},\sigma_p)\rangle\nonumber\\
\hspace{-0.3in}&&+ \,{\rm crossed~ terms}\Big].
\end{eqnarray}
Thus, the total contribution to the S--wave amplitude $\delta
f^{(\gamma)}$ of $\pi^-p$ scattering at threshold, caused by 
one photon in the intermediate state can be written as $\delta f^{(\gamma)} = \sum_X
\delta f^{(X\gamma)}$, where the S--wave amplitude $\delta
f^{(X\gamma)}$ is defined by the reaction $\pi^-p \to X \gamma$.  As
we pointed out in the Introduction, there are strong reasons to
believe that the physical contribution is nearly saturated by the
one--baryon states $X = n$ and $X = \Delta(1232)$.

\section{Contributions  from the 
$\pi^- p \to n \gamma $ and $\pi^- p \to \Delta \gamma $ 
intermediate states}
\subsection{The  $\pi^- p \to n
  \gamma$ contribution}

This case corresponds to $X = n$. It has the physically open radiative
decay channel, which gives rise to an 
 imaginary part of the amplitude. This process is well understood. We will ignore
it in the present discussion of contributions to the real part of the
threshold amplitude.  The contribution is given by the following sum and average 
over intermediate polarization states of the nucleon and photon. 
%
\begin{eqnarray}\label{label11}
  \hspace{-0.3in}&& \Big(1+ \frac{m_{\pi }}{M_N}\Big)\delta f^{(n \gamma)} = 
  \frac{1}{4}\frac{1}{ F^2_{\pi}}\frac{\alpha }{M_N}
  \int \frac{d^3p}{(2\pi)^3 2|\vec{p}|}\sum_{\lambda}\nonumber\\
  \hspace{-0.3in}&&\times {\cal P}
  \int \frac{d^3Q}{(2\pi)^3 2E_n(\vec{Q})}\,\frac{(2\pi )^3 \,
    \delta ^{(3)}(\vec{p} -\vec{Q})}{E_n(\vec{Q}) + |\vec{p}\,| 
    - m_{\pi}- M_p-i0}\nonumber\\
  \hspace{-0.3in}&&\times \, \langle|\langle
  n(\vec{Q},\sigma_n)| e^{\,*}(\vec{p},\lambda)\cdot J^{(-)}_{5}(0)|
  p(\vec{0},\sigma_p)\rangle|^2\,\rangle _{av}\, .
\end{eqnarray}
The matrix element of the axial  current between  nucleon
states is defined by \cite{TE88,MN79}
\begin{eqnarray}\label{label12}
  \hspace{-0.3in} &&\langle n(\vec{Q},\sigma_n)| J^{(-)}_{5\mu}(0)|
  p(\vec{0},\sigma_p)\rangle =\nonumber\\
 \hspace{-0.3in}  &&= g_A\,F_A(\vec{Q}^{\;2})\,\bar{u}_n(\vec{Q},\sigma_n)
\gamma_{\mu}
  \gamma^5 u_p(\vec{0},\sigma_p),
\end{eqnarray}
where $g_A = 1.270(3)$. The axial  form factor
$F_A(\vec{Q}^{\;2})$ can be empirically approximated by 
\begin{eqnarray}\label{label13}
 F_A(\vec{Q}^{\;2}) =  (1 + \vec{Q}^{\;2}/M^2_A)^{-2},
\end{eqnarray}
where $M_A = (960\pm 30)\,{\rm MeV}$ (see e. g. \cite{TE88,MN79}). The squared
matrix element of the axial  current, averaged over
polarizations of the proton and summed over polarizations of the
neutron is
\begin{eqnarray}\label{label14}
  \hspace{-0.3in}  && \sum_{\lambda}\langle |\langle
    n(\vec{Q},\sigma_n)|e^{\,*}(\vec{p},\lambda)\cdot  J^{(-)}_{5}(0)| 
    p(\vec{0},\sigma_p)\rangle|^2\rangle _{av}
  = \nonumber\\
  \hspace{-0.3in}  &&  4 M^2_N
  g^2_A(\vec{Q}^{\,2})\,\frac{1}{2}\sum_{\lambda}{\rm
    tr}\{(\vec{\sigma}\cdot \vec{e}^*(\vec{p},\lambda))(\vec{\sigma}\cdot
  \vec{e}(\vec{p},\lambda))\} \nonumber\\
  \hspace{-0.3in}  &&= 12 M^2_N g^2_A(\vec{Q}^{\,2}),
\end{eqnarray}
where the photon polarization is both transverse and longitudinal.

 Since this is a perturbative contribution to the pion scattering
length, we make for simplicity the heavy--baryon approximation in the following.
The amplitude $\delta f^{(n\gamma)}$ becomes with $p=|\vec{p}|$
\begin{eqnarray}\label{label15}
  \hspace{-0.3in}&&\Big(1+ \frac{m_{\pi}}{M_N}\Big) \delta f^{(n\gamma)} = 
  \nonumber\\
  \hspace{-0.3in}&&\frac{3\alpha}{8\pi^2} 
  \frac{g^2_A}{F^2_{\pi}}
  \, {\cal P}\int^{\infty}_0 \frac{dp\,p\,F^2_A(p^2) }{p -
    m_{\pi}-i0}.
\end{eqnarray}
In the case of the form factor (\ref{label13}) the integral can be
 evaluated analytically. 
We define the  function $I(x)$ with the even {\it (e)} and odd 
{\it (o)} parts 
$I^{(e,o)}(x)=(1/2)(I(x)\pm I(-x))$ as
\begin{eqnarray}\label{label16}
I(x)= {\cal P}\int^{\infty}_0 \frac{dt\,t\, }{t+x-i0}\frac{1}{(1+t^2)^4}.
\end{eqnarray}
Using the notation $d=(1+x^2)$ gives
\begin{eqnarray}\label{label17}
  \hspace{-0.3in}&&I^{(e)}(x) = \frac{\pi}{2}\,\Big(\frac{1}{d^4} -
\frac{1}{2 d^3} - \frac{1}{8 d^2} - \frac{1}{16 d} \Big)\simeq   \nonumber\\
  \hspace{-0.3in}&&\simeq  
  \frac{5\pi}{32}- \frac{35 \pi}{32}\, x^2\,+ \,\ldots
\end{eqnarray}
and 
\begin{eqnarray}\label{label18}
  \hspace{-0.3in}&&I^{(o)}(x) = x\,\Big(\frac{1}{2}\,\frac{\ln x^2}{d^4}
  + \frac{1}{2d^3} +
  \frac{1}{4}\,\frac{1}{d^2} + \frac{1}{6d}\Big) \simeq \nonumber\\  
  \hspace{-0.3in} &&\simeq  x\,\Big(\frac{1}{2}\,\ln x^2\,+\,
  \frac{11}{12}\,+ \,\ldots\Big).
\end{eqnarray}
Using these relations, the amplitude $\delta f^{(n\gamma)}$ in
 Eq.~(\ref{label15}) can be expanded in the small parameter $x~=~x_{\pi}~=
m_{\pi }/M_A$~:
 \begin{eqnarray}\label{label19}
   \hspace{-0.3in}&&\Big(1+ \frac{m_{\pi}}{M_N}\Big) \delta f^{(n\gamma)} =
   \frac{3\alpha}{8\pi^2} \frac{g^2_A}{F^2_{\pi}}M_A I(-x_{\pi}) = \frac{3\alpha }{8\pi ^2}\frac{g^2_A}{F^2_{\pi }} \nonumber 
 \\
   \hspace{-0.3in}&&\times \Big[ \frac{5\pi }{32}M_A
   -  m_{\pi }\Big(\ln \frac{m_{\pi }}{M_A} + \frac{11}{12} + 
{\cal O}({m_{\pi } \over M_A}\Big)
\Big].
\end{eqnarray}
  This result applies 
  identically to the correction of threshold amplitude for the charge
  symmetric reaction $\pi ^+n\to \pi ^+n$ with no change.  The result for
   the scattering lengths for the crossed reactions  $\pi
  ^-n\to\pi ^-n $ and $\pi ^+p\to \pi ^+p$ also follows immediately, since the only change is
  the replacement $m_{\pi }\to -m_{\pi }$ in the 
integral~(\ref{label15}).

The contributions to the  $\pi N$ scattering amplitude at threshold from these
radiative processes with a nucleon are thus 
\begin{eqnarray}\label{label20}
  \hspace{-0.3in}&& \Big(1 + \frac{m_{\pi}}{M_N}\Big)\delta f^{(N \gamma)}= 
  \nonumber\\
  \hspace{-0.3in}&& =  \frac{3\alpha}{8\pi^2} \frac{g^2_A}{F^2_{\pi }} M_A 
  [I^{(e)}(x) t^2_3 + I^{(o)}(x)t_3
  \tau_3]~.
\end{eqnarray}
Here $t_3$ and $\tau_3$ are isospin matrices of pions and nucleons
\footnote {$t_3$ is a diagonal $3\times 3$ matrix with diagonal elements $
  (1,0,-1)$. In the non--diagonal representation it reads
  $(t_3)^{ab}=-i\epsilon ^{ab3}~;~(t_3^2)^{ab}=\delta ^{ab}-\delta
  ^{a3}\delta ^{b3}~$  with $a,b=1,2,3$}.  The amplitude (\ref{label20}) therefore does
not contribute to reactions involving the neutral pion, including
charge exchange.

An alternative derivation is to start from  pseudovector
$\pi NN$ coupling with a Lagrangian term 
\begin{eqnarray}\label{label21} 
{\cal L}_{\pi^-pn}(x) = \frac{g_A}{\sqrt{2}F_{\pi}}\,\bar{n}(x)\gamma^{\mu}\gamma^5
  p(x)\partial_{\mu}\pi^-(x). 
\end{eqnarray}
The result follows replacing $\partial _{\mu }\to \partial _{\mu }
-ie{\cal A}_{\mu }$, where ${\cal A}_{\mu }$ is the e.~m. vector
potential.  This treatment is analogous to that of the matrix element
for the radiative capture process at threshold in the Panofsky ratio.

\subsection{The $\pi^-p \to \Delta
  \gamma$ contribution}

The S--wave amplitude of $\pi^-p$ scattering near threshold has contributions 
from the two intermediate states $\Delta^0\gamma$ and
$\Delta^{++}\gamma$ which appear in the s-- and u--channels of the
reaction $\pi^- + p \to \Delta + \gamma \to \pi^- + p$, respectively. 
The corresponding  two
matrix elements are expressed as those of  the axial currents $\langle
\Delta^0|J^{(-)}_{5\mu}(0)|p\rangle$ and $\langle
\Delta^{++}|J^{(+)}_{5\mu}(0)|p\rangle$.  These are empirically known from
 the
theoretical and experimental analysis of the neutrino production of
the $\Delta^{++}$ resonance \cite{BER65} by the reaction $\nu_{\mu} + p
\to \mu^- + \Delta^{++}$ together with  isospin invariance of the strong low--energy interaction. In the heavy--baryon limit  the matrix
elements  $\langle \Delta^{++}|J^{(+)}_{5\mu}(0)|p\rangle$ and $\langle
\Delta^0|J^{(-)}_{5\mu}(0)|p\rangle$ are defined as follows \cite{BER65}:
\begin{eqnarray}\label{label22} 
  \hspace{-0.3in}  &&\langle \Delta^{++}(\vec{Q},
\sigma_{\Delta})|J^{(+)}_{5\mu}(0)|
  p(\vec{0},\sigma_p)
  \rangle = \nonumber\\
  \hspace{-0.3in} &&= +\,\sqrt{2}\,g_A\,F_A(\vec{Q}^{\,2})\,
\bar{u}_{\Delta^{++}\mu}(\vec{Q},\sigma_{\Delta})u_p(\vec{0},\sigma_p)
\end{eqnarray}
and
\begin{eqnarray}\label{label23} 
  \hspace{-0.3in} &&\langle \Delta^0(\vec{Q},
\sigma_{\Delta})|J^{(-)}_{5\mu}(0)|p(\vec{0},\sigma_p)
  \rangle =\nonumber\\
  \hspace{-0.3in} &&= -\,\sqrt{\frac{2}{3}}\,g_A\,
F_A(\vec{Q}^{\,2})\,\bar{u}_{\Delta^0\mu}(\vec{Q},
\sigma_{\Delta})u_p(\vec{0},\sigma_p).
\end{eqnarray}
Here $u_{\Delta^{++}\mu}(\vec{Q},\sigma_{\Delta})$ and
$u_{\Delta^0\mu}(\vec{Q},\sigma_{\Delta})$ are wave functions of the
$\Delta^{++}$ and $\Delta^0$ resonances \cite{TE88,MN79}.

In the heavy--baryon limit the contributions to the S--wave amplitude
of $\pi^-p$ scattering near threshold from the intermediate
$\Delta^0\gamma$ and $\Delta^{++}\gamma$ states can be expressed using
Eq.~(\ref{label16})
\begin{eqnarray}\label{label24}
\hspace{-0.3in}&&  \Big(1+ \frac{m_{\pi }}{M_N}\Big)
  \delta f^{(\Delta^{++}\gamma)} =
\frac{\alpha}{2\pi^2}\frac{g^2_A}{F^2_{\pi}}M_A
  I(x_{\Delta} +x_{\pi })\nonumber\\
\hspace{-0.3in}&&\Big(1+ \frac{m_{\pi }}{M_N}\Big)\delta f^{(\Delta^0\gamma)} =
  \frac{\alpha}{6\pi^2}\frac{g^2_A}{F^2_{\pi}}M_A
  I(x_{\Delta} -x_{\pi }).
\end{eqnarray}
Here $x_{\Delta } = \omega _{\Delta }/M_A$ with $\omega_{\Delta} =
M_{\Delta} - M_N$. The amplitudes $\delta
f^{(\Delta^{++}\gamma)}$ and $\delta f^{(\Delta^0\gamma)}$ are
calculated with the contributions of transverse and longitudinal
photons.

The same result for the contribution to the S--wave amplitude of
$\pi^-p$ scattering at threshold given by Eq.(\ref{label24}) follows from
 the effective Lagrangian of the $\pi\Delta N$--interactions
\cite{LN71}
\begin{eqnarray}\label{label25} 
  \hspace{-0.3in}&&{\cal L}_{\rm \pi N \Delta}(x)= \frac{g_A}{F_{\pi}}\,\Big[\bar{\Delta}^{++}_{\omega}(x)
  \Theta^{\omega\varphi} p(x) \partial_{\varphi}\pi^+(x) \nonumber\\
  \hspace{-0.3in}&&-\frac{1}{\sqrt{3}}\,\bar{\Delta}^0_{\omega}(x)
  \Theta^{\omega\varphi}
  p(x) \partial_{\varphi}\pi^-(x) + \ldots + {\rm h.c.} \Big],
\end{eqnarray}
where $\Theta^{\omega\varphi} = g^{\,\omega \varphi} - (Z + 1/2)\,
\gamma^{\omega}\gamma^{\varphi}$ and $Z$ is a parameter constrained by
$|Z| \le 1/2$ \cite{LN71}, with the inclusion of the electromagnetic
interaction by the minimal extension:
\begin{eqnarray}\label{label26}
\hspace{-0.3in}&&{\cal L}_{\rm \pi N \Delta \gamma}(x) =
 i\,\frac{e g_A}{F_{\pi}}\Big[\bar{\Delta}^{++}_{\omega}(x)
\Theta^{\omega\varphi} p(x) \pi^+(x)\nonumber\\
\hspace{-0.3in}&&-\,
\frac{1}{\sqrt{3}}\,\bar{\Delta}^0_{\omega}(x)\Theta^{\omega\varphi}
p(x) \pi^-(x) + \ldots \Big] {\cal A}_{\varphi}(x).
\end{eqnarray}
In the heavy--baryon approximation the contribution of the $\Delta
$--resonance to $f^{\Delta \gamma}$ is independent of the
parameter $Z$.

As previously for the case of $N \gamma $ intermediate states this
result readily generalizes to any $\pi ^{\pm }N\to \pi ^{\pm }N$
elastic scattering amplitude.  In the heavy $\Delta $ limit the
contributions are charge symmetric.  For example, the contribution to
the $\pi ^+n $ scattering length is obtained from the $\pi ^-p$ one of
Eqs.~(\ref{label24}) and (\ref{label26}) by the replacements $\Delta ^0\to \Delta ^+$ and $\Delta
^{++}\to \Delta ^-$  with an unchanged
result. In analogy to Eq.(\ref{label20}) we can write this result in
terms of the even and odd functions in the variable $x_{\pi }$:
\[I^{(e,o)}_{\Delta } = [I(x_{\Delta}-x_{\pi})
\pm I(x_{\Delta}+x_{\pi})]/2,\] 
 such that the total contribution from intermediate
$\Delta \gamma$ states is 
\begin{eqnarray}\label{label27}
  \hspace{-0.3in}&&\Big(1 + \frac{m_{\pi }}{M_N}\Big)\delta f^{(\Delta \gamma)} \nonumber
=  \\
  \hspace{-0.3in}&&
\frac{\alpha}{3\pi^2}\,\frac{g^2_A}{F^2_{\pi}}\,M_A \, 
  [2I^{(e)}_{\Delta }t^2_3
  + I^{(o)}_{\Delta }t_3\tau _3].
\end{eqnarray}
We have based our evaluation in Table~1 directly  on this expression. 
While it is possible
to make a perturbative expansion in $x_{\Delta }$
 and/or $x_{\pi }$
 as small parameters, such an expansion
 is poorly convergent and not very informative. This is 
in contrast to Eq. (\ref{label20}) for the nucleon case. 

\section{Discussion }

In this section we investigate numerically the electromagnetic
contributions to the scattering lengths of charged pions 
due to $N \gamma $ and $\Delta \gamma $ intermediate states.
 These intermediate states are associated
with characteristic momenta of order $M_A/2\simeq 500~MeV/c$ or less. The
 results in terms of amplitudes are listed in Table~1, while we discuss the contributions in terms of \% of the typical scale in the text \,\footnote{For numerical analysis we set
  $\omega_{\Delta } = 290\,{\rm MeV}\,;\,\omega_{\Delta } + m_{\pi } =
  430\,{\rm MeV}$ and $\omega_{\Delta } - m_{\pi } = 150\,{\rm MeV}$.
  Note that in the soft--pion limit the mass difference $\omega
  _{\Delta }\neq 0$.}. 

The amplitudes can be expressed in terms of a general classification
of isospin breaking amplitudes at threshold 
\cite{FET01,FET99}

\begin{eqnarray}\label{label28}
  \hspace{-0.3in}&&T^{ba}_{\pi N}(0) = 4\pi\,\Big(1 + \frac{m_{\pi}}{M_N}\Big)
\,f^{ba}_{\pi N}(0) = \nonumber\\
  \hspace{-0.3in}&&= \delta ^{ab}(g_{ba}^++
  \tau ^3g_{ba}^{3+})
  + i\epsilon ^{bac}\tau ^c (g_{ba}^-
  +\tau ^3g_{ba}^{3-}).
\end{eqnarray}
Inside of this classification the terms generated by the present
mechanism by Eqs. (\ref{label19},\ref{label20}) and (\ref{label27}) correspond
predominantly to a large isospin violating contribution proportional
to $t_3^2$, which contributes to the term $g_{ab}^+$. To leading order in the  isospin
breaking amplitude such a term appears also in the heavy baryon ChPT expansion  and has the form \cite{FET01}
\begin{eqnarray}\label{label29}
\delta g_{ba}^+=- 8\pi\, \alpha \,f_1\, (\delta ^{ab} - \delta^{a3}\delta^{b3}),
\end{eqnarray}
where $f_1$ is a ChPT constant. The sign of this term is not known,
but a dimensional estimate gives $F_{\pi }^2|f_1|\simeq M_N/16\pi ^2=6
$ MeV \cite{FET01}; another estimate doubles this value to $12$ MeV \cite{JG02}.  The
amplitudes defined by Eqs. (\ref{label19}) and (\ref{label27}) give a
corresponding term  in the limit $m_{\pi}=0$:
\begin{eqnarray}\label{label30}
\hspace{-0.3in}&&F^2_{\pi}f_1=-\frac{15g^2_A}{ 512\pi}\,M_A\Big(1 + 
\frac{16}{9}\,
\frac{I^{(e)}({\omega _{\Delta }/ M_A})}{I^{(e)}(0)}\Big). 
\end{eqnarray}
This result is not an evaluation of the constant $f_1$ within the
framework of ChPT and EFT, but rather means that we have identified
the main physical mechanism which leads to such a constant.  The
dimensional magnitude estimates quoted above with a large value are
roughly consistent with our contribution of $-9.3$ MeV from the $N\gamma $ channel
alone without the contribution from the $\Delta $ isobar, which gives
a substantial increase.  This isospin-breaking term contributes equally to
each of the 4 elastic $\pi N$ amplitudes. The effect is quite large. We now  discuss  the contributions in more
detail. It is convenient to discuss them in \%
of the experimental $\pi ^-p$ scattering length
 $a_{\pi -p}=0.0883 m_{\pi }^{-1}$~\cite{PSI2}, 
 which sets the scale for the elastic threshold amplitudes
  of  charged pions. 
 In particular, 
while the dimensional  EFT estimate  gives a $\pm 1.4\%$ to $\pm 2.8\%$ contribution 
from $f_1$ to the  level shift in the $\pi ^- $p atom \cite{JG02}, our result of Table 1 gives 
a $+6.1(3)\%$ contribution in the limit $m_{\pi }=0$. 

For physical pions, the intermediate $N\gamma $ state alone
contributes $3.4\%$ to the scattering amplitude, which increases to
$9.4\%$ when the $\Delta $ resonance is included on an equal footing
and degenerate with the nucleon (see Table~1).  The reason for these
surprisingly large numbers is that the present mass scale is larger by
$ M_A/m_{\pi }\simeq 7$ than the chiral one. The e.~m. contribution
based on this term is therefore increased substantially.  The
attractive sign and magnitude  is general,
while the detailed prediction depends on physical assumptions. In
particular, the physical $N\Delta $ mass difference quenches the
$\Delta $ contribution from $5.6\%$ to $2.7\%$.  Any realistic evaluation of the constant
$f_1$ must  include the $\Delta $ isobar
as well as the $N\Delta$ mass splitting.

In the EFT treatment the constant $f_1$ contributes also to the
nucleon e.~m. mass, which is outside the present approach. Even so
this raises an interesting problem.  In our case the corresponding
$f_1$ in the $\pi N$ sector is heavily dominated by the axial
coupling such that it is a magnitude larger than the value  expected
 on the basis of the residual nucleon e.~m. mass term beyond the part due
to the $np$ mass difference.  However, the average nucleon e.~m.
mass in the EFT approach does not depend on $f_1$ but on the sum of
{\it two } ChPT constants $(f_1\,+\,f_3)$\, These are not separate
observables in the nucleon sector~\cite{JG02}.  This suggests that
axial contributions in the two terms may cancel in the sum, at least
to leading order.  A recent study of the e.~m.  corrections in a
heavy--quark model \cite{VL01} indicates that this may indeed be the
case.  The three ChPT constants $(f_1,f_2,f_3)$ linked to e.~m.
effects are explicitly evaluated in this model.  Although not stated
by the authors, the total e.~m.  nucleon mass proportional to
$(f_1+f_3)$ depends only on the nucleon charge and magnetic form
factors in their model.  Their $f_1$ constant, on the other hand, is dominated
by the axial form factor, as we also find.  A massive cancellation in
the sum $f_1+f_3$ eliminates  the axial form factor
from the nucleon e.~m. mass term.  There is then no contradiction
between the chiral result with a contribution to the e.~m.  nucleon
mass, and ours with no contribution to that sector. While the physical
ingredients differ substantially,  our
heavy baryon result is $F_{\pi }^2f_1 = - 25.8(8) $ MeV 
as compared to the heavy quark model value
$-19.5(1.6)$  MeV. 

In addition to this leading contribution our mechanism generates
characteristic terms depending on the non--vanishing pion mass, which
are counterparts to $f_1$.  Once more, such terms give contribution
neither to the charge exchange amplitude nor to the neutral pion
scattering one at threshold. They introduce in particular an
isospin breaking term proportional to $t_3 \tau _3$. In an EFT
expansion such terms have to our knowledge only been considered in
Ref.~\cite{JG02}, where they occur  as a 3rd order term.  
They are outside the leading order isospin breaking considered 
in Refs. ~\cite{FET01,FET99}.  The reason for these
terms is that the pion mass contributes with opposite sign in the
denominators for a direct as compared to a crossed process.

Consider  the case of the $\pi ^-p $ scattering length. 
In the case of  the $n\gamma $ intermediate state,
 the following terms of order $m_{\pi }$ and 
$m_{\pi }\ln m_{\pi }$ from  Eq.~(\ref{label19}) contribute:
\begin{eqnarray}\label{label31}
  \hspace{-0.3in}&&{\cal T}^{(n\gamma )}\equiv 4\pi\,\Big(1+ \frac{m_{\pi}}{M_N}\Big) 
  \delta f^{(n\gamma)}  = \nonumber\\
  \hspace{-0.3in}&& -\,\frac{3\alpha }{2\pi} \, \frac{g^2_A}{F^2_{\pi }}
  \, 
  m_{\pi }\Big(\ln \frac{m_{\pi}}{M_A} + 0.917 + {\cal{O}}({m_{\pi }\over M_A})\Big).
\end{eqnarray}
(for numerics see Table~1, where also the next order terms are included).
This term has an exact counterpart to the same order in Ref.
\cite{JG02} where it is denoted by ${\cal T}^{em}_3(0)$ in the e.~m.
chiral power expansion of the $\pi ^-p$ scattering length (their
Eq.~(9.5): 
\begin{eqnarray}\label{label32}
  \hspace{-0.3in}&&{\cal T}^{em}_3(0) \equiv 4\pi \Big(1+ \frac{m_{\pi}}{M_N}\Big) 
  \delta f^{(em)}_3 = \nonumber\\
  \hspace{-0.3in}&&  - \frac{3\alpha}{2\pi }\frac{g^2_A}{F^2_{\pi }}\, 
  m_{\pi }\Big(\ln\frac{m_{\pi }}{\mu} + 0.891 + {\cal O}(m_{\pi})\Big),
\end{eqnarray}
where $\delta f^{(em)}_3$ is the corresponding contribution to the scattering length.
The leading $m_{\pi}\ln m_{\pi}$ is identical in the 2 cases as
expected and independent of the axial mass $M_A$ and the
dimensional regularization mass $\mu $.  The term proportional to $m_{\pi}$ cannot readily be
compared although the dependence on  $M_A$ and  $\mu $ is weak, since the relation of
these scales to each other is not specified.  Even so it is clear that
the results are reminiscent as to sign, structure and magnitude.
\begin{table*}[htb]
  \caption{Contributions $10^3\,m_{\pi } (\delta f^{(N \gamma }+
 \delta f^{(\Delta \gamma)})$  in the heavy baryon limit to the $\pi N$ 
 scattering lengths corresponding to  
    Eqs.~(\ref{label20})~and~(\ref{label27}). The quoted uncertainty reflects 
the one of  $M_A$.  } 
     
\label{table:1}
\newcommand{\m}{\hphantom{$-$}}
\renewcommand{\arraystretch}{1.5} 
\begin{tabular}{l|lll}
  \hline
$m_{\pi} = 
  0, \omega_{\Delta} = 0$&$(3.0(1)_{N\gamma}
  + 5.3(2)_{\Delta\gamma})\,t^2_3 $&$= 8.3(3)\,t^2_3$ \\
$m_{\pi}
  = 0, \omega_{\Delta} \neq 0$&$(3.0(1)_{N\gamma} + 2.4(1)_{\Delta\gamma})\,t^2_3$ &$=  5.4(2)\,t^2_3$\\
$m_{\pi} 
  \neq 0, \omega_{\Delta} = 0$&$(2.6(1)_{N\gamma} + 4.6(1)_{\Delta\gamma})\,t^2_3$ $+ (- 0.8_{N\gamma} + 
 0.7_{\Delta\gamma})\,t_3\tau_3$&$= 7.2(2)\,t^2_3 + (- 0.1)\,t_3\tau_3$\\
$m_{\pi} 
  \neq 0, \omega_{\Delta} \neq 0$&$(2.6(1)_{N\gamma} + 2.5(1)_{\Delta\gamma})\,t^2_3$ $+ (-0.8_{N\gamma} + 
  0.3_{\Delta\gamma})\,t_3\tau_3$&$= 5.1(2)\,t^2_3 + (-0.5)\,t_3\tau_3$\\
  \hline
\end{tabular}
\end{table*}

The $\Delta $ isobar changes the character of the terms dependent on
the pion mass.  If we  attempt to neglect the $N\Delta $ mass
difference, the value of the isospin breaking isoscalar  amplitude is increased
dramatically by a factor  $25/9$  with
respect to the value for the nucleon only. The isovector contribution on the other hand
becomes negligible  as a consequence of a nearly complete cancellation 
between the nucleon and $\Delta $ terms. 
 This is a mathematical
artifact, however, which does not correspond to the actual physics of
the problem.  Once the $N\Delta$ mass difference is introduced, the $\Delta $ 
term is substantially quenched in both cases.
 As a  consequence  the total
isovector contribution from the joint $ N \gamma $ and $\Delta \gamma $
states  is now a modest  $-\,0.57\%$ net contribution.  This is
nearly half of the  result for the nucleon only
in the absence of the mass splitting.  The $\Delta $ degree of freedom including the $N\Delta $ mass
difference are essential to the understanding of these terms.

Our results are of direct relevance both to the determination of the $\pi $NN
coupling constant as well as to that  of the 
$\sigma_{\pi N}$--term.  In the first case our results indicate that
the e.~m. contribution to the hadronic scattering length combination
$(a_{\pi ^-n}-a_{\pi -p})$, which dominates the GMO dispersion
relation for $g^2_{\pi NN}$ is only $-0.3\%$, such that the evaluation in
Refs. \cite {ERI04,TE03} remains nearly unchanged.  Concerning the
$\sigma_{\pi N}$--term the low--energy isospin symmetric $\pi^c N $
amplitude for charged pions extrapolated to the Cheng--Dashen point
depends linearly on the corresponding isoscalar scattering length.  The
correction for the present e.~m. contribution  diminishes
the value for $\sigma _{\pi N}(2m_{\pi }^2)$ by about $5$ MeV, which
is a sizable fraction of the major correction of $15$ MeV associated
with the further extrapolation to $t=0$ so as to obtain $\sigma _{\pi
  N}(0)$ \cite{SCH04,GAS91}.

It may soon be possible to demonstrate the isospin symmetry
breaking directly for the isospin odd amplitude  in spite
of its small value. This is of particular interest, since it depends
 both on the chiral isospin breaking in the strong sector \cite{FET01,FET99} 
as well as on
 the present breaking for which the dependence
on the form factor is expected to be weak.  It requires a combination
of precision measurements of the 1s level shifts in pionic hydrogen
($a_{\pi ^-p}$) and in pionic deuterium ($a_{\pi ^-p} + a_{\pi ^-n}$) as
well as the charge exchange 1s width in pionic hydrogen ($a_{\pi
  ^-p\to \pi ^0n}$) \cite{PSI1,TE03}.  The first two serve to
eliminate the large isospin breaking in the isoscalar amplitude due to
 $ f_1$,
while the width gives the corresponding charge exchange amplitude.

\section{Conclusion}

We have previously investigated the isospin violating corrections to
the $\pi $N scattering lengths induced by the external Coulomb field
of the extended charge \cite{ERI04}.  These can be well understood in
physical terms to the present level of experimental precision.  Here
we have investigated the intrinsic isospin breaking in the $\pi $N
scattering lengths induced by radiative capture processes with nucleon
and $\Delta $ isobar intermediate states.  The effect is large and
intimately related to p-wave $\pi N$ physics. In view of the obvious
physical origin, there is no reason to believe that it will be
suppressed by systematic cancellations in a more detailed treatment.
The result has no free parameters.  The most important violation
occurs in the isoscalar scattering length for charged pions. This
violating term has the same symmetry property as the one associated
with the ChPT constant $f_1$.  It follows from our result that the
evaluation of this parameter requires that both the nucleon and the
$\Delta $ isobar are taken into account as well as their mass
difference.  In addition, their axial form factor must be included in
accordance with observations.  The finite pion mass generates small
isospin breaking terms, mainly in the isovector amplitude.  The
nucleon gives to leading order in the pion mass a term proportional to
$m_{\pi }\ln \, m_{\pi }$ with a coefficient identical to the one
obtained previously for the $\pi ^-p$ system using EFT \cite{JG02}.
However, the physics of this isospin breaking is governed by the
interplay of the p-wave $\pi N$ contributions from both nucleons and
$\Delta $ isobars and has not been previously investigated.  The
$\Delta \gamma $ intermediate state gives a contribution from the
physical pion mass of opposite sign, which largely cancels the
corresponding nucleon term.  Here the N$\Delta $ mass difference and
the pion mass enter in a non-linear combination.  The result is a
modest net isospin violation in the isovector amplitude.

The importance of our +6.3 \% contribution to the energy shift
 in the 1s state of pionic hydrogen is evident  when compared to  previous results.
Sigg et al.
 \cite{SIG96} find a  $-2.1\pm 0.5\%$ overall correction 
with an  {\it ad hoc}   
estimate of only $-0.7\%$  for the $N\gamma $ contribution  and the sign is  opposite  to ours.
Gasser et al. \cite{JG02} find an  overall  correction of
$-7.2\pm 2.9 \% $ in which the error comes nearly entirely from 
the uncertainty in $f_1$. This error is  estimated  dimensionally
as a one photon loop term, which gives a  magnitude of about half  
of our $f_1$ and of unknown sign.

Our description can be improved beyond the heavy--baryon approximation
along similar lines as here.  We expect a kinematic decrease of the
leading isospin violating term, but no qualitative change.
However, such improvements will give small charge symmetry violating
terms.

Our mechanism for isospin breaking in elastic $\pi N$ scattering gives
contributions to the 1s level shift in pionic hydrogen 30 times larger
than present experimental precision and it is a central feature in the
breaking of isospin symmetry for $\pi N$ scattering at threshold.  So
as to optimally exploit the high precision of ongoing experiments on
pionic hydrogen and deuterium \cite{PSI1,PSI2}, it is therefore
desirable to refine the present results.

\section*{Acknowledgments} 

Torleif Ericson is grateful to Profs. P. Kienle, J. Marton and M.
Faber for their hospitality  and Prof. B. Loiseau for useful discussions.
Andrei Ivanov thanks the TH--division at CERN for hospitality
 during part of this work.
  The present collaboration originated
from a discussion at a CPT$^*$ Workshop at Trento in October of 2003.

\end{document}